\documentclass[letterpaper, 10 pt, conference]{ieeeconf}  

\IEEEoverridecommandlockouts                              

\usepackage{amsfonts, dsfont, mathtools, amsmath, amssymb, bbold, mathrsfs, bm, upgreek}
\usepackage{graphicx, float, epsfig, color, psfrag, adjustbox, enumerate}
\usepackage{subcaption}
\usepackage{subcaption}
\usepackage{refcount, url, cleveref}
\usepackage[noadjust]{cite}
\usepackage[usenames,dvipsnames,svgnames,table]{xcolor}
\usepackage{algorithm}
\usepackage{algpseudocode}

\algrenewcommand\alglinenumber[1]{#1:}
\usepackage{multirow}
\usepackage{multicol}
\usepackage{tikz}
\usetikzlibrary{positioning,arrows}
\usepackage{arydshln}

\title{\LARGE \bf
Control-oriented Modeling of Bend Propagation in an Octopus Arm
}

\author{Tixian Wang$^{1,2}$, Udit Halder$^2$, Ekaterina Gribkova$^3$, Mattia Gazzola$^{1,4,5}$, Prashant G. Mehta$^{1,2}$
\thanks{We gratefully acknowledge financial support from ONR MURI N00014-19-1-2373, NSF/USDA $\#$2019-67021-28989, and NSF EFRI C3 SoRo $\#$1830881. We also acknowledge computing resources provided  by the Blue Waters project (OCI- 0725070, ACI-1238993), a joint effort of the University of Illinois at Urbana-Champaign and its National Center for Supercomputing Applications, and the Extreme Science and Engineering Discovery Environment (XSEDE) Stampede2 (ACI-1548562) at the Texas Advanced Computing Center (TACC) through allocation TG-MCB190004.}%
\thanks{$^{1}$Department of Mechanical Science and Engineering, $^{2}$Coordinated Science Laboratory, $^{3}$Neuroscience Program,
$^{4}$Department of Molecular and Integrative Physiology, $^{5}$National Center for Supercomputing Applications, University of Illinois at Urbana-Champaign. 
  Corresponding e-mail:  {\tt\small udit@illinois.edu}}%
  \thanks{We are thankful to Dr. Rhanor Gillette's lab at the University of Illinois, where the bend propagation video was recorded.}
}
\def\R{{\mathds{R}}}

\def\0{{\mathbb{0}}}
\def\1{{\mathds{1}}}

\def\a{{\mathbf{a}}}
\def\b{{\mathbf{b}}}

\definecolor{db}{RGB}{23,20,119}
\definecolor{dg}{RGB}{2,101,15}

\newtheorem{proposition}{Proposition}[section]

\newtheorem{remark}{Remark}

\usepackage{upgreek}

\newcommand{\set}[1]{\left\{#1\right\}}

\newcommand{\material}[1]{
	\ifthenelse{\equal{#1}{\kappa}}{\upkappa}{
	\ifthenelse{\equal{#1}{\nu}}{\upnu}{
	\ifthenelse{\equal{#1}{\omega}}{\upomega}{
	\ifthenelse{\equal{#1}{\sigma}}{\upsigma}{
	\ifthenelse{\equal{#1}{\theta}}{\uptheta}{
	\mathsf{#1}}}}}}
}

\newcommand{\states}{q}

\renewcommand{\a}{\mathsf{a}}
\renewcommand{\b}{\mathsf{b}}

\newcommand{\muscle}{\text{m}}

\newcommand{\Rmuscle}{\mathsf{R}}
\newcommand{\rigid}{\zeta}
\newcommand{\rigidity}{\mathrm{r}}

\newcommand{\ud}{\,\mathrm{d}}
\newcommand{\diag}{\text{diag}}

\newcommand{\Ve}{\mathcal{V}^e}

\newcommand{\Vr}{\mathcal{V}^{\mathrm{r}}}

\newcommand{\range}{\Big|_{\sigma=0}^s}
\newcommand{\DDl}{\mathcal{D}^\text{linear}}
\newcommand{\DDd}{\mathcal{D}^\text{drag}}
\newcommand{\Dl}{D^\text{linear}}
\newcommand{\Dd}{D^\text{drag}}
\newcommand{\myper}{_\text{per}}
\newcommand{\mytan}{_\text{tan}}
\newcommand{\drag}{ f^\text{drag}}
\newcommand{\rhow}{\rho_\text{water}}
\newcommand{\qr}{\bar{q}}
\newcommand{\pr}{\bar{p}}
\newcommand{\Rho}{\mathrm{P}}
\newcommand{\Angle}{\Theta}
\newcommand{\Htotal}{\mathcal{H}^\text{total}}
\newcommand{\radius}{\gamma^\text{rod}}
\newcommand{\rodtip}{\gamma^\text{tip}}
\newcommand{\rodbase}{\gamma^\text{base}}
\newcommand{\eff }{\text{eff}}

\begin{document}
\bstctlcite{BSTcontrol} 
\maketitle
\thispagestyle{empty}
\pagestyle{empty}


\begin{abstract}
Bend propagation in an octopus arm refers to a stereotypical maneuver whereby an 
octopus pushes a bend (localized region of large curvature) from the base to 
the tip of the arm. 
Bend propagation arises from the complex interplay between mechanics of the 
flexible arm, forces generated by internal muscles, and environmental 
effects (buoyancy and drag) from of the surrounding fluid.  In part due 
to this complexity, much of prior modeling and analysis work has relied 
on the use of high dimensional computational models. 
The contribution of this paper is to present a control-oriented reduced 
order model based upon a novel parametrization of the curvature of 
the octopus arm.  The parametrization is motivated by the experimental 
results. The reduced order model is 
related to and derived from a computational model which is also presented.  The results from the two sets of models are compared using numerical simulations which is shown to lead to useful qualitative insights into bend propagation. A comparison between the reduced order model and experimental data is also reported.          

\end{abstract}



\section{Introduction} \label{sec:intro}

Stereotypical maneuvers of octopus arms has elicited 
much recent interest from biologists and engineers alike~\cite{laschi2012soft, rus2015design, levy2017motor,thuruthel2019emergence,doroudchi2021configuration}.  The most canonical 
of these maneuvers is the {\em bend propagation} where 
the octopus propagates (pushes) the bend from the base of the arm to the tip 
of the arm.  Fig.~\ref{fig:exp} depicts several frames showing the stages of 
a bend propagation maneuver.  The frames are recorded from a freely 
moving octopus in experiments carried out by biologists at the University of Illinois.  The original ground-breaking experiments of bend propagation in octopus arms appeared in a series of papers~\cite{gutfreund1996organization, gutfreund1998patterns, sumbre2001control}. This work was followed by additional papers from the same group that helped clarify the intricate interplay between mechanics of the octopus arm and the environmental effects due to the surrounding fluid~\cite{sumbre2005motor, sumbre2006octopuses,hanassy2015stereotypical}. Apart from octopus arms, bend propagation has also been studied in other biological structures, e.g., in flagella~\cite{brokaw1966bend, brokaw1971bend, hines1978bend}.    


The first numerical investigation of the bend propagation in an octopus arm 
is carried out in a two-part paper~\cite{yekutieli2005dynamic,yekutieli2005dynamic2}.  In 
this seminal work, the flexible arm is modeled computationally as a constrained (constant volume) coupled 
mass spring system.  To accurately capture the physics of bend propagation, 
the model includes the effects of the internal muscle forces 
and external drag forces from surrounding fluid.  Based on
EMG recordings~\cite{gutfreund1998patterns,sumbre2001control}, a 
travelling wave of muscle activation pattern is proposed as control input 
and shown to lead to bend propagation.  

While the computational models are useful to investigate 
the physics of the problem, it is difficult to use these for the purposes of control design~\cite{wang2021optimal, thuruthel2019emergence}.  In part due to the unavailability of suitable control-oriented models, there is no control-theoretic understanding of the traveling wave of muscle 
activation profile (e.g., whether it is optimal in some way).  This motivates the work 
described in this paper which is to develop a control-oriented reduced order model.  To do so, we adopt 
the following two-step approach:  


\begin{figure*}[ht]
	\centering
	\includegraphics[width=2.05\columnwidth,trim = {0pt 0pt 0pt 0pt}]{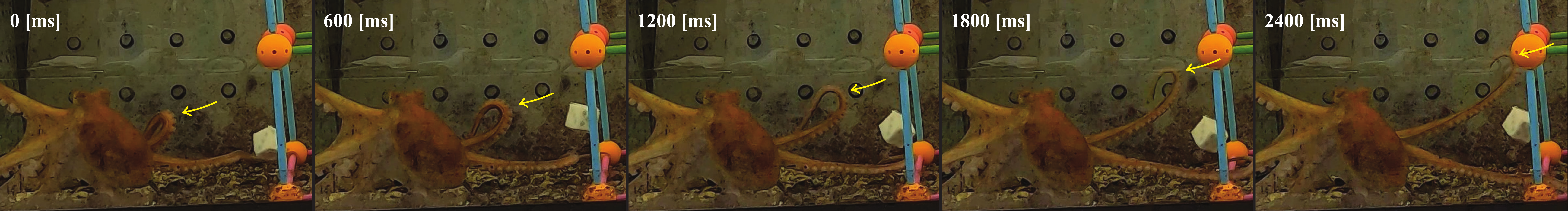}
	\caption{A video sequence of an octopus reaching toward a target. The yellow arrows indicate the location of the bend.}
	\label{fig:exp} 
	\vspace{-10pt}
\end{figure*}

\medskip

\noindent 
\textbf{Step 1: Expressing the computational model as a Hamiltonian control system.} 
The flexible octopus arm with internal muscle forces (control input) is 
modeled using the Cosserat rod theory.  (This approach is closely related to recent papers from our group on modeling and 
control of octopus arms~\cite{chang2020energy,chang2020controlling,wang2021optimal}). 
As in these papers, we express the computational model as a Hamiltonian control system. 
A novel extension in the present paper is to also include a model of non-conservative drag forces (these forces are important to the problem of bend propagation). 
An advantage of expressing the computational model as a 
Hamiltonian control system is that the model reduction can be carried out upon selecting suitable coordinates.  By construction, the reduced order 
model is also a Hamiltonian control system.    

\medskip

\noindent 
\textbf{Step 2: Model reduction.} The reduced order coordinates for bend 
propagation are selected based on the experimental data.  Specifically, the 
curvature of the infinite-dimensional Cosserat rod is parametrized with 
two parameters -- the position of the bend and the expanse of the 
bend.  The resulting reduced order model is a Hamiltonian control 
system with four states.  The control input is due to muscles.    
The effects of drag forces are modeled using the Rayleigh's 
dissipation function.  Related constant and polynomial 
curvature parametrizations for model reduction of soft robots are 
considered in~\cite{della2019control,grazioso2019geometrically,renda2018discrete}.



A detailed numerical comparison is provided between the reduced order and the 
computational model which show that the reduced order model faithfully approximates the bend propagation behavior. The reduced order model is used to obtain regions in the control input space where bend propagation is possible.    
To further investigate these effects, the results from reduced order model are compared against the 
experimental data.  It is shown that the bend propagation crucially 
depends upon matching the wave speed and the experimental 
data shows lower values of the wave speed than the reduced 
order model predictions.

The remainder of this paper is organized as follows.  The computational and reduced order models are presented in Sec.~\ref{sec:full-order} and Sec.~\ref{sec:reduced-order}, respectively.  A comparison of these models appear in Sec.~\ref{sec:numerics} followed by a comparison with experiments in Sec.~\ref{sec:compare_expt}.  The conclusions appear in Sec.~\ref{sec:conclusion}. 



\section{Full Order Model - Cosserat Rod} \label{sec:full-order}


\subsection{Physics of the problem}

Based on the past experimental and numerical work~\cite{sumbre2001control,hanassy2015stereotypical,yekutieli2005dynamic,yekutieli2005dynamic2}, the 
following physical effects are believed to be important for the problem of bend propagation in an octopus arm:   


\medskip

\noindent
\textbf{1) Propagating stiffening wave of muscle activation:} 
The question of {\em how} bend propagates in an octopus arm is not settled.  
Electromyogram (EMG) recordings of muscle activation reveal that the 
arm muscles are activated in unison and that the activation propagates forward 
from the base to the tip of the arm~\cite{gutfreund1998patterns}.  Although 
mathematical modeling of muscles in an octopus arm remains an active 
area of research~\cite{yekutieli2005dynamic, chang2020controlling}, one of the 
effects of muscle activation is to make the arm stiffer.  It has 
been hypothesized that as the stiffening wave travels from the base 
to the tip of the arm, it can cause the bend to propagate~\cite{gutfreund1998patterns}.    

\medskip

\noindent
\textbf{2) Drag effects from the surrounding fluid:}  The question of 
{\em why} octopuses employ bend propagation, e.g., to reach a target in its workspace, remains open.  The difficulty is in part due to 
virtually infinite degrees of freedom associated with a flexible 
octopus arm.  It has been 
hypothesized that bend propagation serves to minimize the 
effective drag from the surrounding fluid~\cite{yekutieli2005dynamic}. With other 
types of reaching motions, e.g., passive unfolding of the 
arm, the effective drag would be greater.   
     
\medskip

Any rigorous modeling effort to understand bend propagation must 
account for these physical effects in addition to inherent elasticity 
of the octopus arm.  For this reason, all of the work on this problem 
has involved development of high-dimensional computational 
models which are of limited use for mathematical analysis and 
control design.  

In the remainder of the present section, we 
describe a novel computational model based on Cosserat rod 
theory.  Because of its interpretation as a Hamiltonian control system (with additional dissipative forces due to drag), it is possible 
to obtain a reduced order model based on a suitable 
parameterization.  



\subsection{Planar Cosserat rod model of an octopus arm}

Since bend propagation has been described to be a planar motion \cite{sumbre2001control}, here we describe a planar Cosserat rod model \cite{antman1995nonlinear, chang2020energy} of an elastic arm. Let $\set{\mathsf{e}_1,\mathsf{e}_2}$ denote a fixed orthonormal basis for the two-dimensional laboratory frame. The independent variables are the time $t\in\R$ and the arc-length $s\in[0,L_0]$ where $L_0$ is the length of the undeformed rod. The subscripts $(\cdot)_t$ and $(\cdot)_s$ denote the partial derivatives with respect to $t$ and $s$, respectively. 

The \textit{state} of the rod is described by the vector-valued function $\states (s,t) = (r(s,t) , \theta(s,t))$  where $r=(x,y)\in\R^2$ is the position vector of the centerline, and the angle $\theta\in[0,2\pi)$ describes the material frame spanned by the orthonormal basis $\set{\mathsf{a}, \mathsf{b}}$, where $\a = \cos \theta \,\mathsf{e}_1 + \sin \theta \, \mathsf{e}_2, ~ \b = -\sin \theta \, \mathsf{e}_1 + \cos \theta \, \mathsf{e}_2$.  The vector $\mathsf{a}$ is defined to be normal to the cross section.  

Let $p(s,t)=\mathcal{M}q_t(s,t)$ denote the momentum where $\mathcal{M}$ is the mass-inertia density matrix.
The kinetic energy $\mathcal{T}$ is given by
\begin{equation}
	\mathcal{T}(p) = \frac{1}{2}\int_0^{L_0} p^\top \mathcal{M}^{-1}p \ud s
\end{equation}
The elastic potential energy of the rod, denoted as $\Ve$, is a functional of the strains: curvature $\kappa$, stretch $\nu_1$ and shear $\nu_2$. The rod's kinematics is given by 
\begin{equation}
	q_s = \left[\begin{array}{c;{2pt/2pt}c}
		Q & 0 \\
		\hdashline[2pt/2pt]
		0 & 1
	\end{array}\right] \begin{bmatrix}
		\nu_1 \\ \nu_2 \\ \kappa
	\end{bmatrix}
\end{equation}
where $Q=\begin{bmatrix}
	\mathsf{a} & \mathsf{b}
\end{bmatrix}$ is the planar rotation matrix.
Let $w=(\nu_1, \nu_2, \kappa)$ denote the triad of deformations. Then we write down the potential energy as
\begin{equation}
	\Ve(q) = \int_0^{L_0} W^e(w(s)) \ud s
\end{equation}
where $W^e:w \mapsto \R$ denotes the elastic stored energy of the rod. Under the linear elasticity settings,  $W^e$ takes the quadratic form as follows:
\begin{equation*}
	W^e(w) = \frac{1}{2}\Big(EA(\nu_1-1)^2 + GA\nu_2^2 + EI\kappa^2 \Big)
\end{equation*}
where $E$ is a constant material Young's modulus and $G$ is the shear modulus; $A$ and $I$ are the rod's cross sectional area and second moment of area, respectively. 

The Hamiltonian $\mathcal{H}$ is the total energy of the rod, i.e., $\mathcal{H}(q,p) = \mathcal{T}(p) + \Ve(q)$ which yields the Hamiltonian control system with muscle actuation, damping, and drag forces. 
\begin{equation}
	\begin{aligned}
	\frac{\ud q}{\ud t} &= \frac{\delta \mathcal{H}}{\delta p} = \frac{\delta \mathcal{T}}{\delta p} \\
	\frac{\ud p}{\ud t} &= -\frac{\delta \mathcal{H}}{\delta q} - c\mathcal{M}^{-1}p + \begin{bmatrix}
	\drag \\ 0
	\end{bmatrix} +  \Rmuscle(q)u
	\end{aligned}
	\label{eq:dynamics-full}
\end{equation}
where $c(s)$ is a damping matrix coefficient that models the inherent viscoelasticity of the rod.
The term $\Rmuscle(q)u$ represents the aggregate forces and couples generated by the musculature where $u(s,t) \in [0,1]$ is the distributed muscle activation function. The modeling of $\Rmuscle$ appears in Sec.~\ref{sec:muscle_full}. The term $\drag$ denotes the forces from the drag model which is described in Sec.~\ref{sec:drag-full}. Note that the drag term contains a zero since the drag does not generate any couple in our simplified model.


\subsection{Muscle model for stiffening the arm}\label{sec:muscle_full}

Physiologically, an octopus arm is composed of a central axial nerve 
cord which is surrounded by densely packed muscle and connective 
tissues. The muscles are of three types: longitudinal, transverse, and 
oblique~\cite{kier2007arrangement}.  In a past work, we have presented mathematical 
models for these muscles based on first principle modeling 
and experimental data~\cite{chang2020controlling}.  

For the purposes of the present paper, motivated in part by the 
stiffening wave hypothesis~\cite{gutfreund1998patterns}, we consider a reduced order model 
whereby the effect of muscle actuation is to simply modify 
the stretch and bending rigidity of the arm
\[
	\left(EA\right)^\eff = E A (1 + \rigid_1 u), \quad \left(EI\right)^\eff = E I (1 + \rigid_2 u)
\] 
where the activation $u\in[0,1]$ and $\rigid_1$, $\rigid_2\geq 0$ are scaling factors -- the effective coefficients for stretch and bend rigidity, respectively. 

A direct consequence of the muscle model is that the muscle internal forces and couples, written compactly as the $\Rmuscle(q) u$ term at the right hand side of equation \eqref{eq:dynamics-full}, arise from a potential function, which we call the `rigidity potential'. We present the result as the following proposition. 
\begin{proposition} \label{prop:rigidity}
	Suppose $u=u(s,t)$ is a given activation of the muscles. Then the muscle-actuated arm is a Hamiltonian control system with the Hamiltonian
	\begin{equation*}
		\Htotal(q,p) = \mathcal{T}(p) + \Ve(q) +  \Vr(q; u) 
	\end{equation*}
	where the muscle rigidity energy is given by
	\begin{equation}
			\Vr = \frac{1}{2}\int_0^{L_0} \Big(\rigid_1EA(s)(\nu_1-1)^2 +\rigid_2EI(s)\kappa^2 \Big)u \ud s
	\label{eq:muscle_rigidity_energy}
	\end{equation}
	In particular, the rigidity forces and couple in~\eqref{eq:dynamics-full} are given by 
	\begin{equation*}
		\Rmuscle(q) u= -\frac{\delta \Vr}{\delta q}
	\end{equation*}
\end{proposition}
\medskip
We omit the proof on account of space. However, the idea of the rigidity potential is very similar to the inherent elasticity of the arm \cite{antman1995nonlinear} or the muscle stored energy function as developed in \cite{chang2020controlling}.

\subsection{Drag model}\label{sec:drag-full}


The drag model is closely based upon~\cite{yekutieli2005dynamic}.  
In the laboratory frame, assuming the shear $\nu_2$ is small,
\begin{equation}
	\drag = -\frac{1}{2}\rhow Q \begin{bmatrix}
		A\mytan c\mytan |v_1|v_1 \\ 
		A\myper c\myper |v_2|v_2
	\end{bmatrix}
\end{equation} 
where $\rhow$ is the density of water, $A\mytan(s)=2\pi \radius(s)$ is 
the surface area of a unit length segment, and $A\myper=2\radius(s)$ is 
the projected area of the unit length segment in the plane perpendicular 
to the normal direction. Here $\radius(s)$ denotes the radius of the circular cross section of the rod.  
The coefficients $c\mytan$ and 
$c\myper$ denote the tangential and perpendicular drag 
coefficients.  Typically, $c\myper$ is much larger 
than $c\mytan$.  Finally, $v_1$ and $v_2$ are the components 
of the velocity $r_t$ in the material frame, i.e., $r_t = v_1\mathsf{a} + v_2\mathsf{b}$.

\subsection{Bend propagation under artificial control}\label{sec:bend-propagation}


\begin{figure}[t]
	\centering
	\includegraphics[width=0.9\columnwidth, trim = {10pt 0pt 0pt 0pt}]{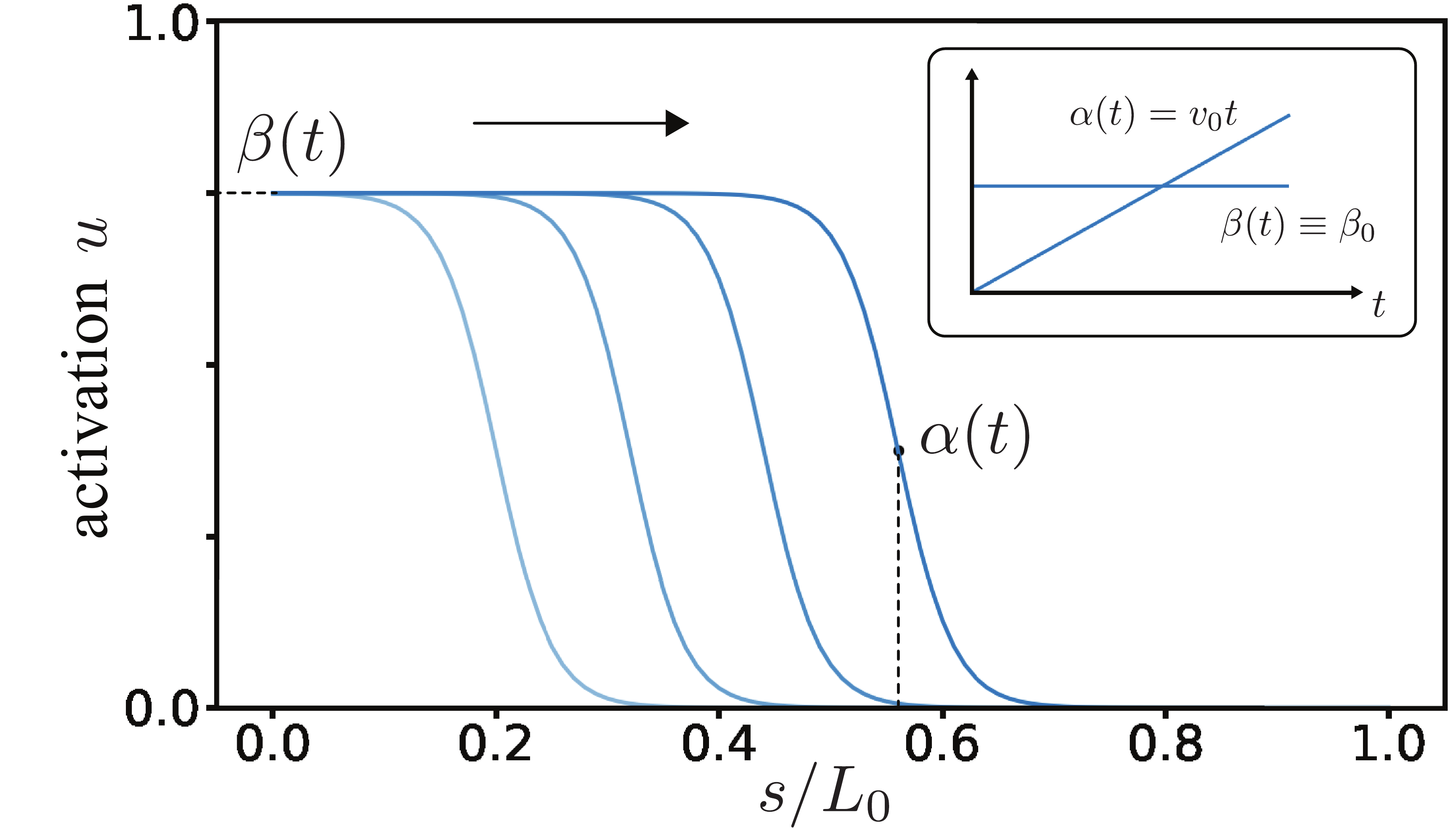}
	\caption{The muscle activation $u$ in the form of a traveling wave function \eqref{eq:stiffening_wave} parameterized by the position of the stiffening wave $\alpha(t)$ and the magnitude of the activation $\beta(t)$. The activation profile shown is a special case where the activation has constant traveling speed $v_0$ and constant magnitude $\beta_0$ (see inset).}
	\label{fig:activation}
	\vspace{-10pt}
\end{figure}

The control problem is to choose an open loop distributed control (activation function) 
$u(s,t)$ for $t\in[0,T]$ and $s\in[0,L_0]$ to push a bend to the tip.  Based on prior 
work~\cite{yekutieli2005dynamic}, we consider a stiffening traveling wave 
\begin{equation}
	u(s,t) = \beta(t)\left(1 - \frac{1}{1 + e^{-200(s - \alpha(t))}}\right)
	\label{eq:stiffening_wave}
\end{equation}
where the control parameters $\alpha(t)\in\R$ determines the traveling 
speed of the activation and $\beta(t)\in[0,1]$ is the magnitude of the 
activation over time (See Fig.~\ref{fig:activation}).

\begin{figure*}[ht]
	\centering
	\includegraphics[width=\textwidth, trim = {0pt 0pt 0pt 0pt}, clip = false]{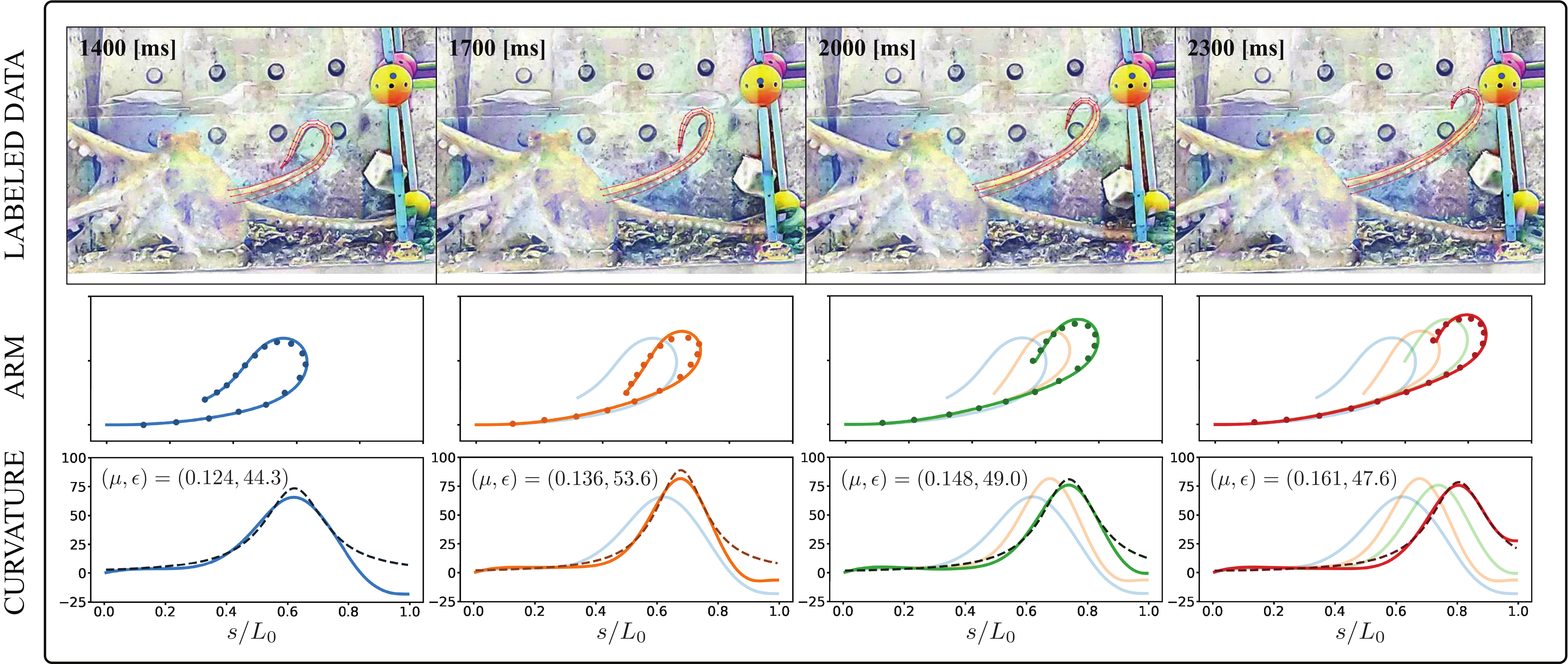}
	\caption{Data labeling and reconstruction from the recorded video: Four frames with labeled data in red are shown on top, indicating the position of the arm. The reconstructed and smoothed arm configurations and curvature profiles are illustrated at the bottom. The dots in the arm configurations are the labeled data points. The dashed curves are the fitted parameterized function $\phi$ in~\eqref{eq:fn-phi} (with $k=1.65$) by minimizing~\eqref{eq:fit-kappa} for the corresponding curvature.}
	\label{fig:reconstruct}
	\vspace{-5pt}
\end{figure*}

\section{Reduced Order Model} \label{sec:reduced-order}

\subsection{Parametrization and its experimental justification}



To construct the reduced-order model, we assume $\nu_1\equiv1$ and $\nu_2\equiv0$ 
and the curvature is of the parametrized form 
\begin{equation}
	\kappa(s,t) = \phi(s; \qr(t))
	\label{eq:fn-kappa}
\end{equation}
where the parameters $\qr=(\mu,\epsilon)\in \R^+ \times \R^+$ and
\begin{equation}
	\phi(s; \mu,\epsilon) = \frac{k\epsilon }{1 + (\epsilon (s-\mu))^2}
	\label{eq:fn-phi}
\end{equation}
where $k>0$ is a fixed gain. We can interpret the coordinates $\mu$ as the center of the bend and $\epsilon$ as the `extent' of the bend, i.e., the higher $\epsilon$ is, the sharper the bend is.

The justification for the proposed ansatz is presented with the aid of 
Fig.~\ref{fig:reconstruct}.  As shown in the bottom panel of the figure, 
the ansatz accurately models the experimentally observed curvature 
profiles in the bend propagation.  The details of experiments, video 
capture system, data acquisition, labeling, reconstruction, and 
smoothing are outside the scope of this paper. Related work can be found in \cite{yekutieli2007analyzing, kim2021physics}.



\subsection{State of the rod}

In terms of the reduced order coordinates $\qr(t)$, the 
state of an inextensible, un-shearable rod is obtained as
\begin{equation}
	\begin{aligned}
		\theta(s,t) &= \theta_0 + \int_0^s \kappa(\sigma,t)\ud \sigma =: \theta_0 + \Angle(s; \qr(t))\\
		r(s,t) &= \int_0^s \begin{bmatrix}
			\cos\theta(\sigma,t) \\ \sin\theta(\sigma,t)
		\end{bmatrix}\ud \sigma =: \Rho(s; \qr(t)) 
	\end{aligned}
	\label{eq:pos-orien}
\end{equation}
where $\theta_0$ is a given constant, denoting the orientation of the rod at the base. 
The velocity and angular velocity are expressed as
\begin{equation*}
	\begin{aligned}
		&\begin{aligned}
			r_t(s,t) &= \Rho_t(s;\qr(t)) = \Rho_{\mu}(s; \qr(t))\dot{\mu}(t) + \Rho_{\epsilon}(s; \qr(t))\dot{\epsilon}(t) 
		\end{aligned} \\
		&\theta_t(s,t) = \Angle_t(s;\qr(t))= \Angle_{\mu}(s; \qr(t))\dot{\mu}(t) + \Angle_{\epsilon}(s; \qr(t))\dot{\epsilon}(t) 
	\end{aligned}
\end{equation*}

From Proposition~\ref{prop:rigidity}, 
the Cosserat rod model with muscle forces is a Hamiltonian control 
system.  Our objective is to express the Hamilton control system in terms 
of the reduced order coordinates $(\mu,\epsilon)$.  This involves expressing 
the kinetic and potential energy of the elastic rod in terms 
of $(\mu,\epsilon)$ and their time derivatives $(\dot{\mu},\dot{\epsilon})$ 
(Sec.~\ref{sec:kinetic}).  This is followed by modeling of the 
change in the potential energy due to muscle stiffening (Sec.~\ref{sec:muscle}).     
The drag force is not conservative and the reduced order model for 
the same is derived by expressing the Rayleigh dissipation function (Sec.~\ref{sec:drag-reduced}).  Based on these calculations, the reduced order model is obtained from writing the Lagrangian and the Hamiltonian (Sec.~\ref{sec:dynamics-reduced}).  

\begin{figure*}[!t]
	\centering
	\includegraphics[width=\textwidth, trim = {0pt 0pt 0pt 0pt}]{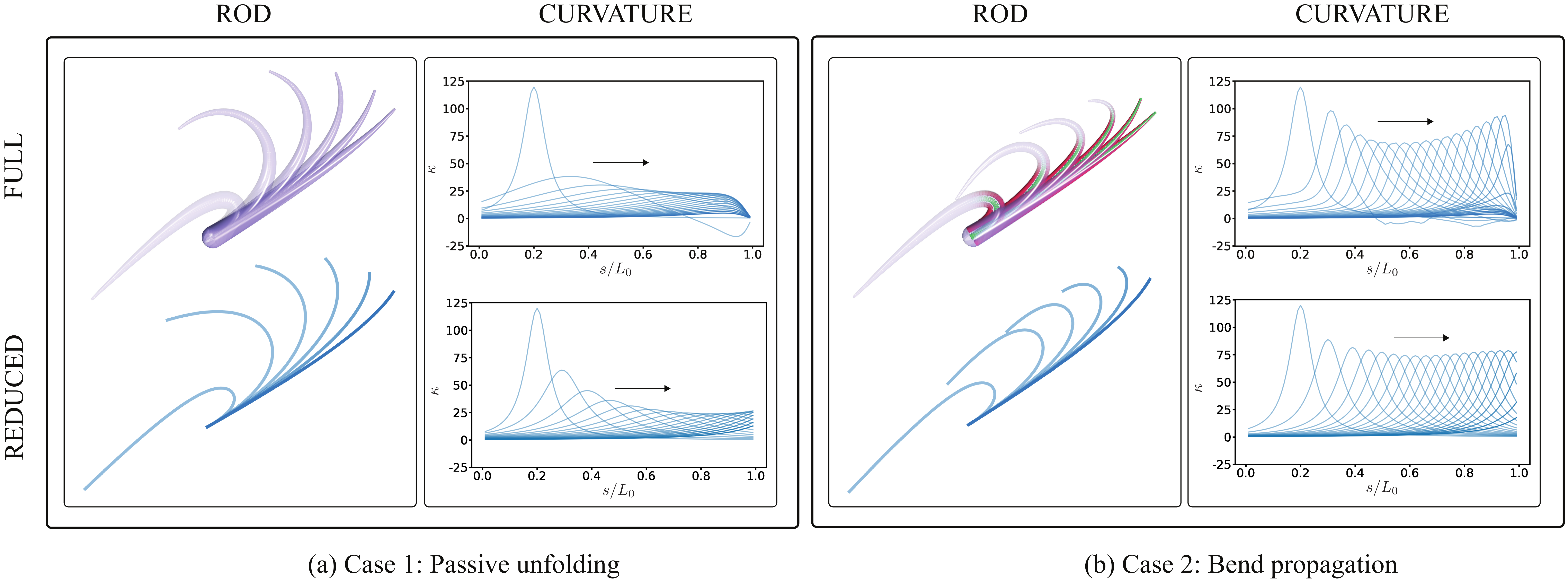}
	\caption{Comparison of rod and curvature profiles between full order model and reduced order model in two cases: passive unfolding and bend propagation. We select six time instances for each rod configurations. The full order model simulation is shown in faded purple rods with the muscle structures illustrated in red for longitudinal muscles and green for the transverse muscle. The reduced order model simulation is shown in fade-in blue curves. A sequence of time snapshots are shown in blue for curvature $\kappa(s,t)$ and the time interval is 0.25 [s] for case 1 and 0.05 [s] for case 2. The black arrow indicates the wave traveling direction. In case 1, the simulation time is 5 [s] and reduced order model simulation stops at 4.6 [s]. In case 2, the simulation time is 1.25 [s] and reduced order model simulation stops at 1.1 [s].}
	\label{fig:compare-rod}
	\vspace{-5pt}
\end{figure*}

\subsection{Potential and kinetic energy of the rod} \label{sec:kinetic}
The rod is inextensible and un-shearable.  
Using the linear elastic model, its potential energy is  
\begin{equation}
	\begin{aligned}
		\Ve 
		= \frac{1}{2}\int_0^{L_0} EI(s)\phi^2(s;\qr) \ud s =: V^e(\qr)
	\end{aligned}
\end{equation}

The kinetic energy is expressed as
\begin{equation}
	T(\qr,\dot{\qr}) = \frac{1}{2}\dot{\qr}^\top M(\qr) \dot{\qr} \nonumber
\end{equation}
where the mass matrix $M(\qr)=\left[\begin{smallmatrix}
	M_1 & M_2 \\ M_2 & M_3
\end{smallmatrix}\right]\in\R^{2\times2}$ whose entries are given by
\begin{equation}
	\begin{aligned}
		M_1 &= \int_0^{L_0} \rho A\Rho_{\mu}^\top\Rho_{\mu} + \rho I \Angle_{\mu}^2 \ud s \\
		M_2 &= \int_0^{L_0} \rho A\Rho_{\mu}^\top\Rho_{\epsilon} + \rho I \Angle_{\mu}\Angle_{\epsilon} \ud s \\
		M_3 &= \int_0^{L_0} \rho A\Rho_{\epsilon}^\top\Rho_{\epsilon} + \rho I \Angle_{\epsilon}^2 \ud s \\
	\end{aligned}
	\label{eq:mass-2}
\end{equation}


\subsection{Potential energy of the muscle} \label{sec:muscle}


The reduced order description of muscle potential energy is entirely analogous to the potential 
energy of the elastic rod.  Suppose $\{u(s,t):0\leq s\leq L_0,t\geq 0\}$ is a given muscle activation; then
\begin{equation}
\hspace*{-1pt}
	\begin{aligned}
		\Vr 
		&= \frac{1}{2}\int_0^{L_0} \rigid_2EI(s)\phi^2(s;\qr) u(s,t) \ud s =: V^\rigidity(\qr; u)
	\end{aligned}
	\label{eq:muscle_rigidity_energy_reduced}
\end{equation}


\subsection{Dissipation modeling} \label{sec:drag-reduced}

The drag force is an example of nonlinear damping.  It is readily modeled using   
Rayleigh's dissipation function (see~\cite{goldstein2002classical}):
\begin{equation*}
	\mathcal{D} = \int_0^{L_0} \xi(s)|v|^n \ud s
\end{equation*}
with $n=3$.  Here $\xi(s)$ is a given dissipation coefficient.  The choice $n=2$ is 
an example of linear damping.


\medskip

\noindent
\textbf{Drag:} The dissipation function
\begin{equation}
	\begin{aligned}
		\DDd &= \int_0^{L_0} \frac{1}{3} \Big(c_1(s)|v_1|^3 + c_2(s) |v_2|^3 \Big) \ud s \\
		&= \int_0^{L_0} \frac{1}{3} \Big(c_1(s)|\mathrm{v}_1(s;\qr,\dot{\qr})|^3 + c_2(s) |\mathrm{v}_2(s;\qr,\dot{\qr})|^3 \Big) \ud s \\
		&=: \Dd(\qr, \dot{\qr})
	\end{aligned}
\end{equation}
where $c_1(s) = \frac{1}{2}\rhow A\mytan(s)c\mytan$ and $c_2= \frac{1}{2}\rhow A\myper(s)c\myper$ 
are the drag coefficients for tangential and perpendicular directions, respectively.
The velocity in the material frame in terms of state $\qr$ is given by 
\begin{equation*}
	\begin{aligned}
	\begin{bmatrix}
			\mathrm{v}_1 \\ \mathrm{v}_2
		\end{bmatrix} := \begin{bmatrix}
			\cos(\theta_0+\Angle) & \sin(\theta_0+\Angle) \\
			-\sin(\theta_0+\Angle) & \cos(\theta_0+\Angle)
		\end{bmatrix} \Rho_t
	\end{aligned}		
\end{equation*}



\medskip

\noindent
\textbf{Linear damping:} Assuming the damping matrix of the form $c(s) = \text{diag}(c_3(s), c_3(s), c_4(s))$, where $c_3(s)$ and $c_4(s)$ are the translation and rotation damping coefficients, respectively; the dissipation function for linear damping becomes
\begin{equation*}
	\begin{aligned}
		\DDl &= \int_0^{L_0} \frac{1}{2}c_3(s)(x_t^2 + y_t^2) + \frac{1}{2}c_4(s)\theta_t^2 \ud s \\
		&= \int_0^{L_0} \frac{1}{2}c_3(s)\Rho_t(s;\qr)^\top\Rho_t(s;\qr) + \frac{1}{2}c_4(s)\Angle_t^2(s;\qr) \ud s \\
		&=: \Dl(\qr, \dot{\qr})
	\end{aligned}
\end{equation*}
\begin{figure}[!t]
	\centering
	\includegraphics[width=0.9\columnwidth, trim = {20pt 0pt 0pt 0pt}]{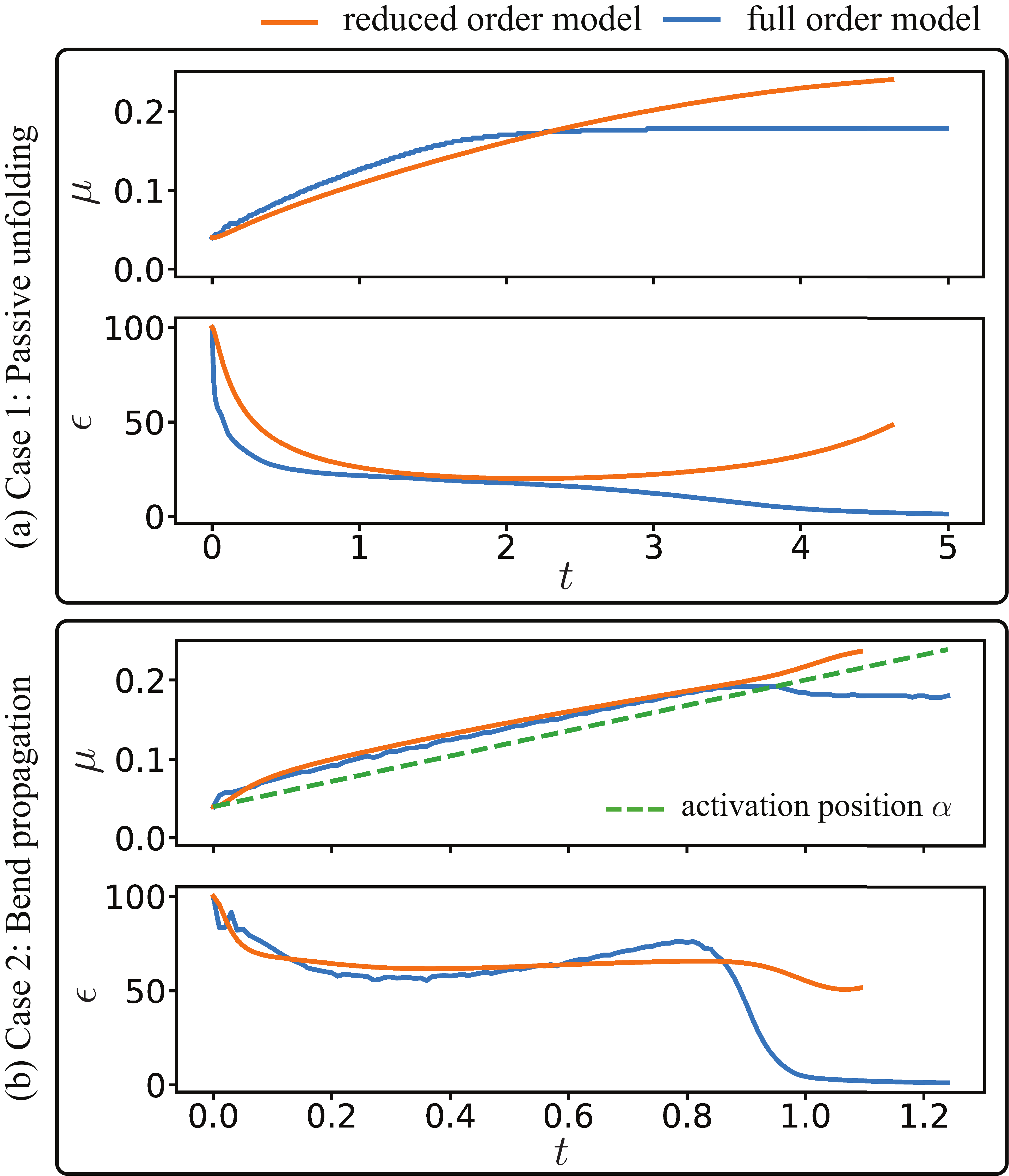}
	\caption{Comparison of curvature coordinates $\left(\mu(t),\epsilon(t)\right)$ between full order model and reduced order model. In bend propagation case, the additional dashed line in green indicates the position of the activation $\alpha(t)$.
	}
	\label{fig:compare-states}
	\vspace{-5pt}
\end{figure}

\begin{figure}[t]
	\centering
	\includegraphics[width=1.0\columnwidth, trim = {0pt 0pt 0pt 0pt}]{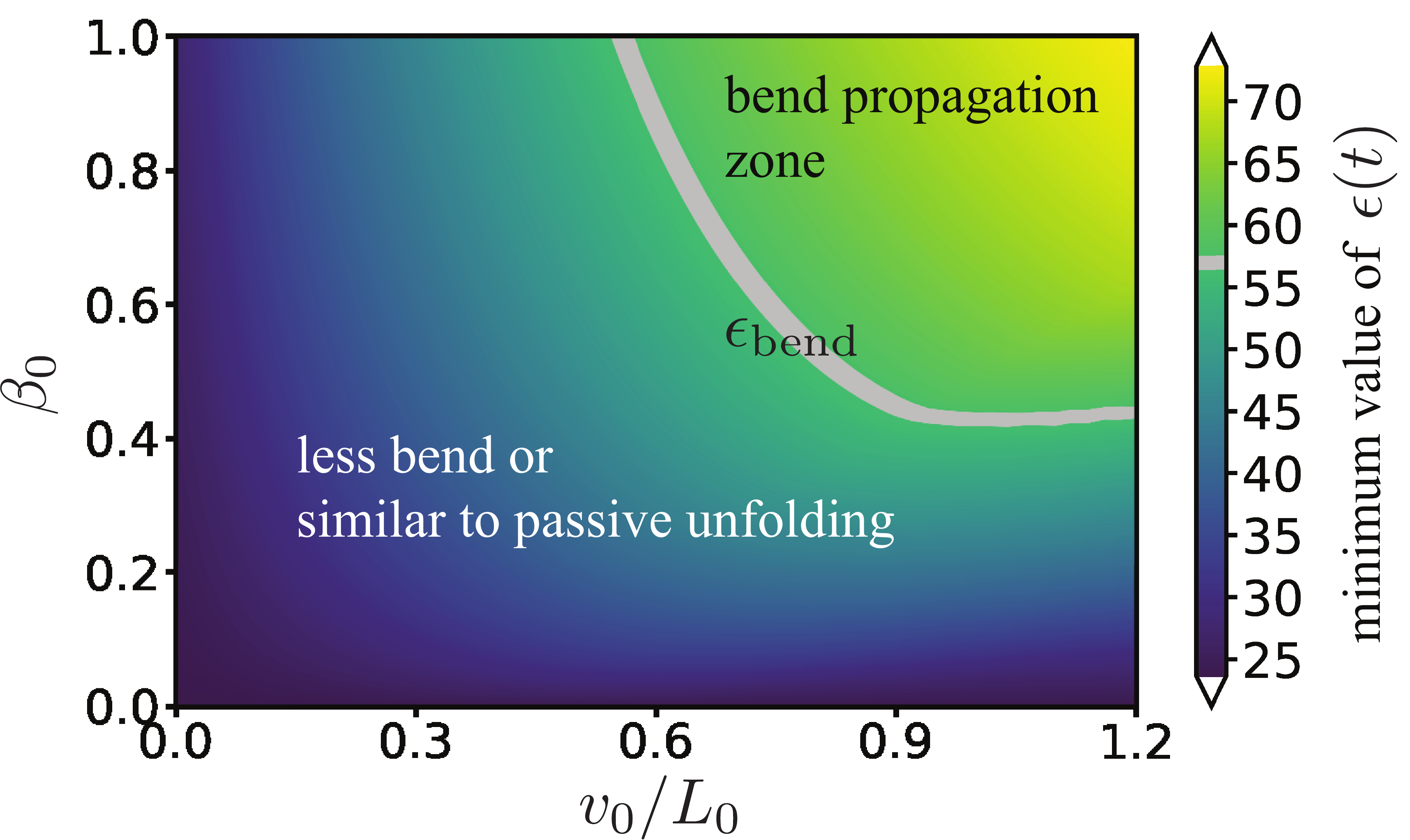}
	\caption{The minimum value of the curvature coordinate $\epsilon(t)$ under different control parameters, i.e., activation speed $v_0/L_0$ and magnitude $\beta_0$.}
	\label{fig:control-param}
	\vspace{-10pt}
\end{figure}

\subsection{Reduced order model} \label{sec:dynamics-reduced}
The Lagrangian 
$
	L(\qr, \dot{\qr}) := T(\qr, \dot{\qr}) - V^e(\qr) - V^\rigidity(\qr)
$
and in the presence of dissipative forces, the E-L equations are
\[
\frac{\ud}{\ud t}\frac{\partial L}{\partial \dot{\qr}} - \frac{\partial L}{\partial \qr}  = -\frac{\partial \Dl}{\partial \dot{\qr}} - \frac{\partial \Dd}{\partial \dot{\qr}}
\]

It is also straightforward to write the Hamiltonian control system.  Define the momentum as $\pr=M(\qr)\dot{\qr}$, so the kinetic energy is $\tilde{T}(\qr, \pr)=\frac{1}{2}\pr^\top M^{-1}(\qr) \pr$. The Hamiltonian of the reduced order system is $H(\qr, \pr)=\tilde{T}(\qr,\pr) + V^e(\qr)$. Then the reduced order model is 
\begin{equation}
	\begin{aligned}
		\dot{\qr} &= M^{-1}(\qr) \pr \\
		\dot{\pr} &= -\frac{\partial \tilde{T}}{\partial \qr} - \frac{\partial V^e}{\partial \qr} -\frac{\partial \Dl}{\partial \dot{\qr}} - \frac{\partial \Dd}{\partial \dot{\qr}} + F(u^\muscle)
	\end{aligned}\label{eq:rom}
\end{equation}
and $F(u^\muscle)=-\frac{\partial V^\rigidity}{\partial \qr} $ is the control 
input on account of muscles.  
\begin{table}[!h]
	\centering
	\caption{Parameters for models and numeric simulation}
	\begin{tabular}{ccc}
		\hline
		\hline\noalign{\smallskip}
		Parameter & Description & Numerical value \\
		\hline\noalign{\smallskip}
		\multicolumn{3}{c}{{\bf Rod model}}\\
		$L_0$ & length of the undeformed rod [cm] & $20$ \\
		$\rodbase$ & rod base radius [cm] & $1$ \\
		$\rodtip$ & rod tip radius [cm] & $0.1$ \\
		$\rho$ & density [kg/${\text{m}}^3$] & $1042$ \\
		$c(s)$ & damping matrix [kg/s]  & $0.01I$ for all $s$ \\
		$E$ & Young's modulus [kPa] & $10$ \\
		\hline\noalign{\smallskip}
		\multicolumn{3}{c}{{\bf Drag model}}\\
		$\rhow$ & water density [kg/${\text{m}}^3$] & $1022$ \\
		$c\mytan$ & tangential drag coefficient & $0.155$ \\
		$c\myper$ & normal drag coefficient & $5.065$ \\
		\hline\noalign{\smallskip}
		\multicolumn{3}{c}{{\bf Muscle model}}\\
		$\rigid_{1,2}$ & effective coefficients & $15$ \\
		\hline\noalign{\smallskip}
		\multicolumn{3}{c}{{\bf Numerics}}\\
		$\Delta t$ & Discrete time step-size [s] & $10^{-5}$ (full) \\
		 &  & $10^{-3}$ (reduced) \\
		$N$ & number of discrete segments & $100$ \\
		\hline 
	\end{tabular}
	\label{tab:num_para}
	\vspace*{-10pt}
\end{table}

The Hamiltonian control system~\eqref{eq:rom} is the reduced order model of the 
computational model~\eqref{eq:dynamics-full}.  Explicit forms of the dynamic equations appear in Appendix~\ref{appdx:explicit}. 
\medskip

\begin{remark}
Based on experimental data and its numerical comparison (see Fig.~\ref{fig:reconstruct}), 
the form~\eqref{eq:fn-phi} of the curvature with its two parameters 
$(\mu,\epsilon)$ adequately captures the bend propagation.  The major 
limitation of the model is that it is applicable only to inextensible rods.  
An octopus arm is known for its hyper-extensibility.  Although some of the 
earlier studies considered inextensible models~\cite{yekutieli2005dynamic}, more recent 
experiments suggest that extensibility may be important to the 
bend propagation~\cite{hanassy2015stereotypical}.  
Therefore, enhancements of this basic model to include extensibility effects is an important next step.     

\end{remark}

\section{Simulation results} \label{sec:numerics}

\begin{figure*}[!t]
	\centering
	\includegraphics[width=\textwidth, trim = {0pt 0pt 0pt 0pt}, clip = false]{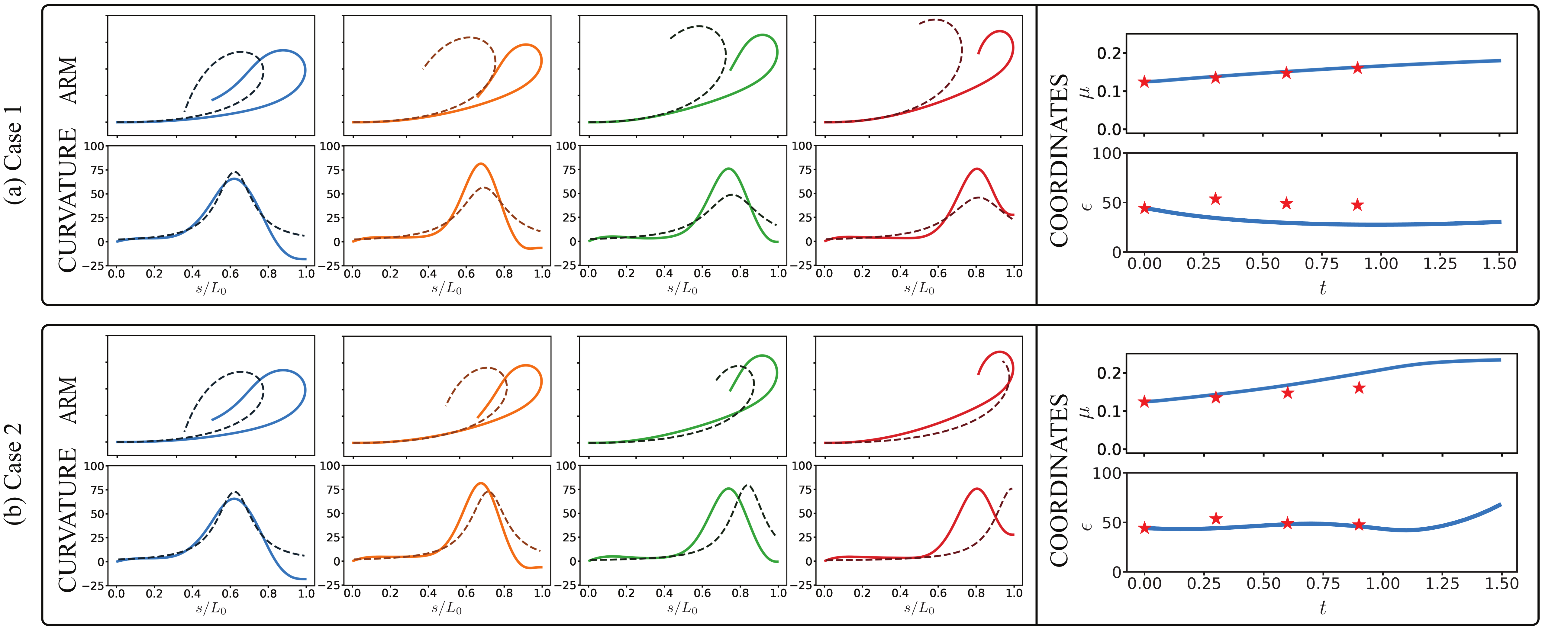}
	\caption{Comparison of the reduced order model with the experimental data: Two cases with different control parameters are presented. In each panel, the first two rows show the arm configurations and curvatures at time instances when the experimental data is available. The experimental data is shown in solid colored curves and the simulation results of the reduced order model are shown in dashed curves. On the right column, we represent the coordinates $(\mu, \epsilon)$. Experimental data points are shown in red stars.}
	\label{fig:comparison_expt}
	\vspace{-5pt}
\end{figure*}

In this section, we demonstrate a comparison between the full order model and the reduced order model with the numerical results for two cases: passive unfolding and bend propagation. The full order model dynamics are solved using the existing software tool~\textit{Elastica}~\cite{gazzola2018forward, zhang2019modeling}.  
The reduced order model is simulated using a direct numerical discretization of the fourth order ODE. 
In all simulations, the variable $\radius(s) = \rodtip s/L_0 + \rodbase(1 - s/L_0)$ gives the radius profile of a tapered rod.  The mass-inertia density matrix is given by $\mathcal{M}=\diag(\rho A, \rho A, \rho I)$. $\rho$ is the material density of the rod, $A=\pi(\radius)^2$ and $I=\frac{A^2}{4\pi}$ are the cross section area and second moment of area, respectively. The shear modulus is given by $G=\frac{4}{3}\cdot \frac{E}{2(1+\text{Poisson's ratio})}$~\cite{gazzola2018forward} where we take the Poisson's ratio to be 0.5 by assuming a perfectly incompressible isotropic material.
The simulation parameters for both the full order model and reduced order model are tabulated in Table~\ref{tab:num_para}.  The reduced order model uses the same parameters as full order model except that the discrete time step-size is smaller by a factor of 100. To avoid singularity of the mass matrix $M(\qr)$~\eqref{eq:mass-2}, we set a threshold such that the reduced order model simulation stops when $\det(M(\qr))\leq1\times10^{-16}$.

\subsection{Comparison of two models} \label{subsec:cases}

For the reduced order model, we use~\eqref{eq:fn-phi} as the parameterized function for curvature where we set the constant gain $k=1.2$. The base orientation $\theta_0=\frac{\pi}{6}$ is kept fixed in both the models. Both rods are initialized with curvature according to~\eqref{eq:fn-phi} using $\mu(0)=0.2L_0$ and $\epsilon(0)=20/L_0$ so that the rod start with a bent configuration. 



%
%
%

\subsubsection{Passive unfolding}

The first case is a simple passive unfolding movement of the rod without any muscle actuation. As shown in Fig.~\ref{fig:compare-rod}(a), both rods unfold due to the inherent elasticity and end up in the rest configuration, which is a straight rod. The curvature decays over time and the position of the peak curvature shifts from the initial bend to the tip.

\subsubsection{Bend propagation}

The second case is designed to reproduce the bend propagation in both full and reduced order models. Both models use the same activation function~\eqref{eq:stiffening_wave} with magnitude $\beta(t)\equiv 0.5$ and $\alpha(t)=(0.2+0.8t)L_0$ which means the activation travels with a given constant speed. As depicted in Fig.~\ref{fig:compare-rod}(b), the bend is formed initially close to the base of the rod and then propagates along the arm until the rod becomes straight. The curvature profiles show the traveling wave of curvature from the initial bend towards the tip. Unlike the unfolding case, the magnitude of the curvature decreases at the beginning but then is maintained at a near constant value and even increases a little bit in the end. Both the forward and reduced order model have the similar peak curvature around 75 [m$^{-1}$].

Fig.~\ref{fig:compare-states} depicts a quantitative comparison between the full order model and the reduced order model by illustrating the trajectories of the selected coordinates of the curvatures, $\mu$ and $\epsilon$. For the reduced order model, these two coordinates are directly obtained from the system state. For the full order model, we find the least square solution $\phi(s;\mu,\epsilon)$ given the original curvature $\kappa(s)$, i.e., we solve the following minimization problem using a gradient descent algorithm:
\begin{eqnarray}
	\underset{\mu, \epsilon}{\min}\ \int_0^{L_0} \left|\kappa(s) - \phi(s;\mu, \epsilon)\right|^2 \ud s
	\label{eq:fit-kappa}
\end{eqnarray}
The comparison of the coordinates in Fig.~\ref{fig:compare-states} indicates that the reduced order model serves as a good approximation to the full order model, especially the center of the bend $\mu(t)$ in the bend propagation case in the reduced order model matches the full model well. Fig.~\ref{fig:compare-states}(b) shows that the trajectory of $\mu(t)$ is slightly ahead of $\alpha(t)$ which indicates that the activation as a stiffening wave pushes the bend forward from the initial bend to the tip.

\subsection{Control parameters of the reduced order model}


As mentioned in Sec.~\ref{sec:bend-propagation}, we are interested in two control parameters of the muscle activation, the traveling speed and the magnitude of the stiffening wave. In this section, we assume the activation profile of $u$ as in \eqref{eq:stiffening_wave} with $\beta(t)\equiv \beta_0$ and $\alpha(t)=v_0 t$, i.e. the wave propagates with a constant speed $v_0$ and with a constant magnitude $\beta_0$. The evolution of the rod in the reduced order model is studied by varying these two control parameters.


In order to qualitatively describe the rod behavior, we use the minimum value of $\epsilon(t)$ over the simulation time horizon, indicating how well the bend is kept during the movement. The higher value of this quantity means that the rod keeps the bend or the activation even makes the bend sharper, and hence is classified as bend propagation movement. On the other hand, the lower value indicates that the rod loses the bend and behaves more like the unfolding movement. A critical value $\epsilon_{\text{bend}}$ separates these two behaviors.
The color map of the minimum value of $\epsilon(t)$ versus varying activation speed $v_0/L_0$ and magnitude $\beta_0$ is shown in Fig.~\ref{fig:control-param}. The grey curve of the critical value $\epsilon_{\text{bend}}=57\pm0.5$ separates the map into two parts: top right is the bend propagation zone while bottom left corresponds to the rod behavior with less bend or similar to unfolding.

%

\section{Comparison with the experimental data} \label{sec:compare_expt}
In this section, we provide a comparison between the reduced order model and the experimental data.  In the reduced order model, the states are initialized according to the curvature of the arm in the first frame of the experiment, i.e. with $k = 1.65$ and $(\mu, \epsilon) = (0.124,44.3)$ (see Fig.~\ref{fig:reconstruct}).  The parameters are the same as the ones used in the preceding section with two modifications made to obtain a reasonable match: the drag coefficients $c\mytan=0.62, c\myper=20.26$ are larger by a factor of 4 and the base orientation $\theta_0 = 0$.
The comparison with the experiment is elucidated with the aid of two cases depicted in Fig.~\ref{fig:comparison_expt}. 

\medskip
\noindent \textbf{Case 1:} We use the following activation: $\alpha(t)=0.04t$ and $\beta(t) \equiv0.2$. As seen from Fig.~\ref{fig:comparison_expt}(a), with these values of the activation, the reduced order model matches the $\mu$ coordinate of the experiment well, i.e., the curvature moves with the same speed. However, unlike the experiment, the rod physically does not exhibit bend propagation behavior as the curvature decays over time.  With this choice of wave speed, the reduced order model shows the characteristic of passive unfolding.   

\medskip

\noindent \textbf{Case 2:} To obtain bend propagation in the reduced order model, we increased the activation to $\alpha(t) = 0.16t$ and $\beta(t) \equiv1.0$.  The results are depicted in Fig.~\ref{fig:comparison_expt}(b). In this case, although the rod exhibits bend propagation with the reduced order model, the speed of the bend propagation is faster than the speed in the experimental data. 

\remark{
While these preliminary results are encouraging, bend propagation in real octopuses likely has a number of other contributing factors, such as differences in mechanical properties of the flexible arm, added mass effects due to surrounding fluid, and possibly others. Additionally, bell-shaped bend velocity profiles have been reported in literature~\cite{gutfreund1998patterns, sumbre2001control} which suggests different types of control inputs.  The control-oriented reduced order model sets the stage for both investigating some of these effects as well as optimizing the control input to better match the experimental data.  This is a subject of continuing work.   



}

\section{Conclusion and Future Work} \label{sec:conclusion}

In this paper, we investigate the bend propagation behavior in an octopus arm using two models: an elastic arm modeled as a planar Cosserat rod, and a control-oriented reduced order model that captures the characteristics of bend propagation by parameterizing the curvature function and introducing the coordinates as new states. Effects of muscle actuation and drag are considered in the system dynamics of both models. Muscles are modeled to effectively increase the rigidity of the arm upon actuation. In both the models, a stiffening wave of muscle actuation is demonstrated to push an initial bend towards the tip, thus achieving the bend propagation movement.
Comparisons through numerical experiments show that the reduced order model is a good approximation to the full order model and has great potential for control problems related to bend propagation. 
In this work, a first attempt is made to compare experimental data with our control-oriented model. A comprehensive treatment of the problem would require more rigorous experimental data analysis and solving  an optimization problem of control parameters. The current reduced order model can also be extended to a more general form by adding a parameterized function for the stretch so that we can also consider the elongation of the arm during bend propagation.

\bibliographystyle{IEEEtran}
\bibliography{reference}

\appendices
\renewcommand{\thelemma}{A-\arabic{section}.\arabic{lemma}}
\renewcommand{\thetheorem}{A-\arabic{section}.\arabic{theorem}}
\renewcommand{\theequation}{A-\arabic{equation}}
\renewcommand{\thedefinition}{A-\arabic{definition}}
\setcounter{lemma}{0}
\setcounter{theorem}{0}
\setcounter{equation}{0}

\section{Explicit calculations} \label{appdx:explicit}
Note that all the subscripts in this Appendix denote the partial derivatives.  The subscript $(\cdot)_{\qr_i}$ refers to the partial derivatives with respect to either $\mu$ or $\epsilon$.
\subsection{Rod model}
\begin{equation}
	\begin{aligned}
		\kappa(s,t) &= \phi(s;\qr(t)) = \frac{k\epsilon(t)}{1 + (\epsilon(t)(s-\mu(t)))^2} \\
		\theta(s,t)-\theta_0 &= \Angle(s;\qr(t)) = k\arctan\left(\epsilon(t)(\sigma-\mu(t))\right)\range \\
	\end{aligned} \nonumber
\end{equation}
\noindent
\textbf{First-order partial derivatives}
\begin{equation}
	\begin{aligned}
		\phi_\mu &= \phi^2\cdot \frac{2\epsilon(s-\mu)}{k},\quad \phi_\epsilon = \frac{\phi}{\epsilon} - \phi^2\cdot \frac{2(s-\mu)^2}{k} \\
		\Angle_\mu &= -\phi(\sigma; \qr)\range, \quad \Angle_\epsilon = \phi(\sigma;\qr)\cdot \frac{\sigma-\mu}{\epsilon}\range \\
		\Rho_{\qr_i} &= \int_0^s \begin{bmatrix}
			-\sin(\theta_0+\Angle(\sigma; \qr)) \\ \cos(\theta_0+\Angle(\sigma;\qr))
		\end{bmatrix} \Angle_{\qr_i}(\sigma; \qr) \ud \sigma
	\end{aligned} \nonumber
\end{equation}
\noindent
\textbf{Second-order partial derivatives}
\begin{equation*}
	\begin{aligned}
		\Angle_{\mu\mu} &= -\phi_\mu(\sigma;\mu,\epsilon)\range,\quad \Angle_{\mu\epsilon} = -\phi_\epsilon(\sigma;\mu,\epsilon)\range,\quad \\
		\Angle_{\epsilon\mu} &= \Big(\phi_\mu(\sigma;\mu,\epsilon)\cdot \frac{\sigma-\mu}{\epsilon} -\frac{ \phi(\sigma;\mu,\epsilon)}{\epsilon}\Big)\range \\
		\Angle_{\epsilon\epsilon} &= \Big(\phi_\epsilon(\sigma;\mu,\epsilon)\cdot\frac{\sigma-\mu}{\epsilon} - \phi(\sigma;\mu,\epsilon)\cdot\frac{\sigma-\mu}{\epsilon^2}\Big)\range
	\end{aligned}
\end{equation*}
and \begin{equation}
	\Rho_{\qr_i\qr_j}= \int_0^s Q \begin{bmatrix}
			-\Angle_{\qr_i}\Angle_{\qr_j} \\ \Angle_{\qr_i\qr_j}
		\end{bmatrix} \ud \sigma \nonumber
\end{equation}
Recall that $Q$ is the planar rotation matrix.

\subsection{Kinetic energy}
In order to evaluate the Coriolis and centrifugal terms, $\dot{M}(\qr)\dot{\qr}$ and $-\frac{\partial T}{\partial \qr}$, we need to calculate the following terms:
\begin{equation}
	\frac{\partial M}{\partial \qr_i} = \begin{bmatrix}
		M_{1\qr_i} & M_{2\qr_i} \\ M_{2\qr_i} & M_{3\qr_i} 
	\end{bmatrix} \nonumber
\end{equation}
where
\begin{equation}
	\begin{aligned}
		M_{1\qr_i} &= \int_0^{L_0}  2\rho A \Rho_\mu^\top\Rho_{\mu\qr_i}  + 2\rho I \Angle_\mu \Angle_{\mu\qr_i} \ud s \\
		M_{2\qr_i}  &= \begin{aligned} \int_0^{L_0} & \rho A (\Rho_{\mu\qr_i}^\top\Rho_\epsilon + \Rho_\mu^\top\Rho_{\epsilon\qr_i}) + \\
		&\rho I (\Angle_{\mu\qr_i}\Angle_\epsilon + \Angle_\mu\Angle_{\epsilon\qr_i}) \ud s \end{aligned} \\
		M_{3\qr_i} &= \int_0^{L_0}  2\rho A \Rho_\epsilon^\top \Rho_{\epsilon\qr_i} + 2\rho I \Angle_\epsilon\Angle_{\epsilon\qr_i}\ud s 
	\end{aligned} \nonumber
\end{equation}

\subsection{Total potential energy}
The explicit form of the restoring forces and couples from the elastic potential energy plus the rigidity forces and couples from the muscle rigidity is given by
\begin{equation}
	\frac{\partial V^e}{\partial \qr} + \frac{\partial V^\rigidity}{\partial \qr} = \int_0^{L_0} EI(s)(1 + \zeta_2u)\phi(s;\qr)\begin{bmatrix}
		\phi_\mu(s;\qr) \\ \phi_\epsilon(s;\qr)
	\end{bmatrix} \ud s \nonumber
\end{equation}

\subsection{Dissipation and drag}
The explicit drag force is given by 
\begin{equation}
	\frac{\partial \Dd}{\partial \dot{\qr}} = \int_0^{L_0} \begin{bmatrix}
		\Rho_\mu^\top \\ \Rho_\epsilon^\top
	\end{bmatrix} Q \begin{bmatrix}
		c_1(s)|\mathrm{v}_1|\mathrm{v}_1 \\ c_2(s)|\mathrm{v}_2|\mathrm{v}_2
	\end{bmatrix}  \ud s \nonumber
\end{equation}
and explicit linear damping is 
\begin{equation}
	\frac{\partial \Dl}{\partial \dot{\qr}} = \int_0^{L_0} \left( c_3(s)\begin{bmatrix}
		\Rho_\mu^\top \\ \Rho_\epsilon^\top
	\end{bmatrix} \Rho_t + c_4(s)\begin{bmatrix}
		\Angle_\mu \\ \Angle_\epsilon
	\end{bmatrix} \Angle_t \right) \ud s \nonumber
\end{equation}

\end{document}